\documentstyle{crnart12}
\def\beq{\begin{equation}}
\def\eeq{\end{equation}}
\def\bea{\begin{eqnarray}}
\def\eea{\end{eqnarray}}
\def\bq{\begin{quote}}
\def\eq{\end{quote}}

\def\gappeq{\mathrel{\rlap {\raise.5ex\hbox{$>$}}
{\lower.5ex\hbox{$\sim$}}}}

\def\lappeq{\mathrel{\rlap{\raise.5ex\hbox{$<$}}
{\lower.5ex\hbox{$\sim$}}}}
\begin{document}
\pagestyle{empty}
\begin{flushright}
{CERN-TH.6721/92}\\
{ACT-22/92}\\
{CTP-TAMU-75/92}\\
{UMN-TH-1117/92}\\
\end{flushright}
\begin{center}
{\bf FLIPPED HEAVY NEUTRINOS:}\\
{\bf  FROM THE SOLAR NEUTRINO PROBLEM TO BARYOGENESIS}\\
\vspace*{1cm} {\bf John Ellis} \\
\vspace*{0.3cm}
{\it Theoretical Physics Division, CERN} \\
{\it CH - 1211 Geneva 23} \\
\vspace*{0.3cm}
and \\
\vspace*{0.3cm}
{\bf D.V. Nanopoulos}\\
{\it Center for Theoretical
Physics, Department of Physics, Texas A\&M University}\\
{\it College Station, TX 77843-4242, USA}\\
 {\it and}\\
 {\it Astroparticle Physics
Group, Houston Advanced Research Center (HARC)}\\
{\it The Woodlands, TX 77382, USA}\\
\vspace*{0.3cm}
and \\
\vspace*{0.3cm}
 {\bf Keith A. Olive} \\
\vspace{0.3cm}
 {\it School of Physics and Astronomy, University of Minnesota} \\
{\it Minneapolis, MN 55455, USA}\\
\vspace*{0.5cm}
{\bf ABSTRACT} \\ \end{center}
\noindent
We discuss baryogenesis using the flipped $SU(5)$ model
 for lepton mass matrices. We show that the generalized see-saw mechanism
in this model can not only provide MSW neutrino mixing suitable for
solving the solar neutrino problem, and supply a hot dark matter
 candidate ($\nu_\tau$) with mass $0(10)eV$ as indicated by recent
COBE results, but can also naturally account for the
baryon asymmetry of the universe.  Heavy singlet neutrino decay
generates a net lepton asymmetry which is subsequently
reprocessed  by nonperturbative
electroweak interactions. We evaluate the baryon asymmetry so produced
in light of the constraints
that the COBE observations put on inflationary
cosmologies, finding it comfortably consistent with observation.

\vspace*{0.3cm}

\begin{flushleft}
CERN-TH.6721/92 \\
{ACT-22/92}\\
{CTP-TAMU-75/92}\\
UMN-TH-1117/92 \\
November 1992
\end{flushleft}
\vfill\eject
\vfill\eject
\pagestyle{empty}
\clearpage\mbox{}\clearpage
\pagestyle{plain}
\setcounter{page}{1}
 {\newcommand{\la}{\mbox{\raisebox{-.6ex}{$\stackrel{<}{\sim}$}}}
{\newcommand{\ga}{\mbox{\raisebox{-.6ex}{$\stackrel{>}{\sim}$}}}

Cosmology, astrophysics and particle physics have for many years been
in constant and fruitful interaction. Much of this was triggered by the
advent of Grand Unified Theories (GUTs), which predicted novel phenomena
that had the potential to solve many of the most important problems in
cosmology, and were perhaps easier to test in astrophysics than in the
laboratory. Characteristic examples of these phenomena were the
baryon-number-violating interactions that could have been responsible
for the baryon-to-entropy ratio $n_B/s$, and the
lepton-number-violating interactions leading to neutrino masses that
could help explain the apparent solar neutrino deficit and the Dark
Matter in the Universe.

The initial euphoria in astroparticle physics was somewhat
dampened when these GUT miracles were subjected to closer
inspection. For example, it has been realized that non-perturbative
effects in the Standard Model violate baryon  and lepton number
conservation \cite{krs1}, could have erased a GUT-generated baryon
 asymmetry~\cite{arnmac}, and
might even be able to generate the baryon-to entropy ratio themselves~\cite{S}.
Moreover, for some time the solar neutrino story did not settle
down~\cite{bah},
and hot neutrino Dark Matter became disfavoured by theorists of galaxy
formation~\cite{white}.

The balance of optimism has now been somewhat redressed.
The realization is increasing that non-perturbative
baryon-number-violating interactions have a limited ability to erase a
GUT baryon or lepton asymmetry~\cite{krs2,cdeo3}, and as yet no demonstrated
ability to generate a baryon-to-entropy ratio of the observed
magnitude~\cite{S}
Moreover, recent observations of the solar neutrinos seem to confirm that there
is indeed a solar neutrino deficit that is difficult to explain by
astrophysics alone~\cite{gall}, and the recent COBE~\cite{Cobe} observations of
large-scale density perturbations tend to favour models with a mixture
of hot and cold Dark Matter.

The closest approximation to the traditional GUT framework that is
permitted by the usual string model-building techniques is flipped
$SU(5)$~\cite{barr,ant1}. We are therefore motivated to revisit these familiar
baryon- and lepton-number-violating astroparticle interfaces in the context of
this model. Indeed, two of us (J.E and D.V.N.) together with J.
Lopez~\cite{eln}
pointed out recently that flipped $SU(5)$ ties together in a natural way
the Mikheyev-Smirnov-Wolfenstein (MSW)~\cite{msw} interpretation of the
apparent
solar neutrino deficit and the O(10) eV neutrino favoured by COBE and
other measurements of large-scale structure, and suggests that
$\nu_{\mu}$ - $\nu_{\tau}$ oscillations could be observable in
forthcoming accelerator neutrino experiments. In this paper we extend
the previous analysis to include an elegant mechanism for baryogenesis,
modelled on the non-GUT heavy neutrino decay mechanism utilizing
sphaleron interactions discussed by Fukugita and Yanagida~\cite{fy1}.

The central role in our analysis is played by the massive singlet
neutrino that is an unavoidable facet of flipped $SU(5)$. It leads to a
see-saw neutrino mass matrix~\cite{ant1,ant2,leon,ln} that accommodates
naturall
solar, COBE and other data. We argue here that its decays naturally provide a
lepton asymmetry which non-perturbative Standard Model interactions
recycle into the observed baryon-to-entropy ratio. Heavy neutrino
decays had been discussed previously~\cite{dstuff} as a possible source of the
baryon asymmetry, but before the realization of the important implications of
Standard Model baryon-number-violating interactions.  Also predating this
realization were previous analyses of baryogenesis in flipped
$SU(5)$~\cite{cehno2}, which were, frankly, rather complicated and yielded
asymmetries of marginal magnitude.
 Here we demonstrate explicitly, using our
previous flipped neutrino analysis and the latest weak interaction
parameters~\cite{eln}, and allowing for reheating of the Universe after
inflation at the scale indicated by the COBE density perturbations~\cite{cdo2},
that decays of the massive flipped neutrino can explain naturally the
observed baryon-to-entropy ratio.

Before starting our analysis, we first remind the reader of the
superpotential of the minimal flipped $SU(5)$ model, and introduce our
notation. The renormalizable terms in the superpotential are~\cite{ant1}
\begin{equation}
W = {\lambda_1}_{ij}F_iF_jh + {\lambda_2}_{ij}F_i{\bar f}_j\bar h +
{\lambda_3}_{ij}\bar f_il^c_jh + \lambda_4 HHh + \lambda_5 \bar H \bar H \bar h
   \\
+ {\lambda_6}_{ij}F_i\bar
H \phi_j  +  {\lambda_7}_i h \bar h \phi_i  +  {\lambda_8}_{ijk}
\phi_i \phi_j \phi_k
\label{i}
\end{equation}
where the $F_i$, $\bar f_i$ are {\bf 10}- and ${\bf {\bar 5}}$-dimensional
matte
representations of SU(5), the $H$ and $\bar H$ are {\bf {10}} and ${\bf \bar
{10
   }}$~
representations, the $h$ and $\bar h$ are {\bf {5}} and ${\bf {\bar 5}}$ Higgs
representations, and the charged $l^c_i$ and neutral $\phi_m$ are
singlets of SU(5). The couplings $\lambda_{1,2,3}$ give masses to the
charge -1/3 quarks, charge 2/3 quarks and charged leptons respectively,
the couplings $\lambda_{4,5}$ separate naturally the light doublet and
heavy triplet components of the {\bf {5}} and ${\bf {\bar 5}}$ Higgs
multiplets,
the coupling $\lambda_6$ plays an important role in the neutrino mass
matrix as we shall see shortly, and the couplings $\lambda_{7,8}$ need
not concern us here. In addition, the neutrino mass matrix involves a
crucial non-renormalizable $F_i F_j H H$ coupling
${\lambda_9}_{ij}$~\cite{ant2}
   ,
 as we
shall also see shortly.

Prior to the realization of the importance of non-perturbative
electroweak  interactions in connection with the baryon asymmetry, we described
several possibilities for the production of a net baryon number in the context
of flipped $SU(5)$~\cite{cehno2}.
 The alternatives depended on 1)
the field content and 2) reheating subsequent to inflation.  In the
minimal flipped $SU(5)$ model~\cite{ant1}, there are a limited number of fields
whose decays can be used to produce a net baryon number.  This has
been the chief obstacle to writing down a satisfactory model of
baryogenesis. Moreover, the breaking of flipped $SU(5)$ is accompanied
by entropy production~\cite{cehno1} which further impedes a large asymmetry,
though we show below that the entropy production was previously overestimated.

Examining the field content of the flipped $SU(5)$
model above, the most obvious candidates to produce a baryon
asymmetry would be the decays of color triplets in the $h$ and $\bar h$
multiplets.
But it is well known~\cite{nw} that one can not generate an asymmetry with a
single $h$, $\bar h$ pair.  Moreover, it is not possible to add
additional Higgs representations to this model without defeating the
natural doublet-triplet splitting mechanism~\cite{cehno2}:  additional $h$,
$\bar h$ fields would require additional $H$, $\bar H$ fields, and these would
give rise to light ``uneaten" baryon-number-violating $u_H$, $d_H$
fields after GUT symmetry breaking, unless further modifications to
the model are employed.  Here, we will stick to the minimal model.
As an alternative, one might also consider the massive $\nu_{H}^{c}$,
$\nu_{\bar H}^{c}$ combination orthogonal to the
flaton $\Phi$, discussed below, as a possible
source of baryon asymmetry, but the decay of such a field (which is
through the $h$ or $\bar h$ field) is always CP-conserving.
We were therefore only left with the massive $\phi_m$ fields,
which combine with $\nu_{i}^{c}$ to form massive Dirac states
through the see-saw process.
The only possibility for a net baryon number, therefore, appeared~\cite{cehno2}
to be the four-body decay channel $\phi \rightarrow FFF\bar f$.  As
this decay channel is heavily suppressed relative to the dominant
two-body channels, a sizeable asymmetry was not a hopeful
prospect.

Moreover, the final baryon asymmetry in any model with an ``intermediate"
scale is in general suppressed if the gauge symmetry is broken along a
flat direction~\cite{Y}. In flipped $SU(5)$, $SU(5) \times U(1)$ is broken down
to $SU(3) \times SU(2) \times U(1)$ by the F- and D- flat direction
$\langle H \rangle = \langle\bar H\rangle \not= 0$.  Because of the
flatness, which is removed only by supersymmetry breaking, the
gauge is symmetry is broken late by the flaton $\Phi$,
the linear combination of $\nu_{H}^{c}$ and $\nu_{\bar H}^{c}$ in $H$
and $\bar H$ that develops a vev, and this is generally accompanied by
the production of entropy~\cite{cehno1}.  Typically, the avoidance of excessive
entropy production
sets stringent constraints on the vev associated with
the flat direction~\cite{eeno1}.

Inflation plays an important role in that the amount of
entropy produced by flaton decays depends heavily on whether or
not the gauge symmetry is restored subsequent to inflationary
reheating.  Previously we found that a sufficient baryon asymmetry
could only be produced in the minimal flipped $SU(5)$ model in the strong
reheating scenario~\cite{cehno2}, i.e. the one in which the gauge symmetry was
restored.  Even then the outcome was only marginal.  In the weak
reheating scenario, which as we emphasize below is the more
natural possibility, the branching ratio to the four-body $\Delta B
\not= 0$ channel was far too small to be significant~\cite{cehno2}.

In what follows, we present a simple mechanism to generate the
baryon asymmetry based on the idea of Fukugita and Yanagida~\cite{fy1},
which generates a lepton asymmetry that is then recycled
by sphaleron effects to produce $\Delta B \not= 0$.  Indeed, the
previous scenario $(\phi \rightarrow FFF\bar f)$ generates no net $B-
L$, in which case sphaleron effects wipe out any $\Delta B = \Delta L
\not= 0$\footnote{Unless some of the lepton asymmetry can be stored in a weakly
coupled field such as $e_R$~\cite{cdeo3}.}.

The generation of a net lepton asymmetry in flipped $SU(5)$ is
closely related to the see-saw mechanism generating neutrino
masses.  Thus there may be a close connection between the solar
neutrino problem and the baryon asymmetry.  From the flipped
superpotential above (\ref{i}) one can easily write down the neutrino
mass matrix~\cite{eln}
\begin{equation}
( \nu_i, \nu^c_i, \phi_i)
\hskip 0.5cm
\left( \begin{array}{ccc}
    0 & m_u & 0 \\
     m_u & {{\lambda_9}_{ij}}{V^2 \over M_{nr}} &  {{\lambda_6}_{ij}}V \\
       0 & {{\lambda_6}_{ij}}V & \mu_{ij} \end{array} \right)
\left( \begin{array}{c}
\nu_i \\ \nu^c_i \\ \phi_j \end{array} \right)
\end{equation}
 where $m_u = {{\lambda_2}_{ij}} v$ is
the up-quark mass matrix, $V=\langle H \rangle
= \langle \overline {\!H} \rangle \sim 10^{15}GeV$ is the vacuum expectation
value of the Higgs ${\bf 10}$ and $\bf{\overline {10}}$ breaking $SU(5)
\times U(1)$, $\mu \simeq \lambda_8 \langle \phi_0
\rangle \simeq10^{17}GeV$, where $\langle\phi_0 \rangle$ is the
vev of one of the singlet fields, and is responsible for giving large
masses to the other singlets.  (Note there is still a singlet combination
with a small vev which provides the necessary $h,~\bar h$ mixing.)
 Finally, we include a mass term from a nonrenormalizable
piece of the superpotential~\cite{ant2}
\begin{equation}
{\cal L}~\ni~{{\lambda_9}_{ij}}~\frac{1}{M_{nr}}
{}~F_i~F_j~\bar H~\bar H~\rightarrow~{\lambda_{qij}}~ {V^2 \over M_{nr}}
\nu_i^c~\nu_j^c
\end{equation}
Noting the hierarchy $m_u~\ll~\lambda_9 \frac {V^{2}}
{M_{nr}} \ll V \ll \mu$, we can identify the principal mass
eigenvalues and eigenstates.  There are states which are
predominantly $\phi_{i}$ with masses $\mu$, and there is a set of
states which are predominantly $\nu_{i}^{c}$ with masses
${{{\lambda^2}_6}_{ij}} \frac {V^{2}} {\mu}$.  The light neutrino masses are
just $m_\nu ~\simeq  \frac  {m_{u}^{2} \mu} {(\lambda_6^2~V^2)}$ (when
$\lambda_9~\mu~\ll~\lambda_{6}^{2} M_{nr}$):
\begin{equation}
(m_{\phi +\ldots})_{ij} \simeq \mu_{ij}~;\quad (m_{\nu^c +\ldots})_{ij} \simeq
\lambda^2_{6ij} \frac{V^2}{\mu}~,\quad (m_{\nu +\ldots})_{ij} \simeq
\bigg(\frac{m^2_u\mu}{\lambda^2_6V^2}\bigg)_{ij}
\label{III}
\end{equation}
Given the additional hierarchy
in the up-quark mass matrix ${{\lambda_2}_{1}}$ : ${{\lambda_2}_{2}}$ :
${{\lambda_2}_{3}} = {10^{- 4}}$ : $10^{-2}$ : $1$, and $\lambda_6 \simeq {1
\over 3}$, one finds for the three light neutrino masses
$({{m_\nu}_e},~{{m_\nu}_\mu},~{{m_\nu}_\tau}) \sim
 (10^{-7} eV,~10^{-3} eV,~10eV)$.
 Furthermore, the expected mixing between
 $\nu_{e}$ and $\nu_{\mu}$ is
 $\sin^{2}2\theta_{e\mu} \sim \frac {m_{u}}{m_{c}}
{}~\sim10^{-2}$ for
$\Delta m^{2} \sim 10^{-6}$.  The fact that
${{m_\nu}_\tau}\sim 10eV$ implies that this pattern of
neutrino masses can simultaneously provide cosmological hot dark
matter with $\Omega_{HDM}  = {{\Omega_\nu}_\tau} \simeq 0.3$, as
suggested by the COBE and other observations of large-scale structure,
and give
$\nu_e$ - $\nu_{\mu}$ mixing sufficient to solve the solar neutrino
problem \cite{eln}.
We now show that the {\em same} neutrino mass matrix can also be used
to account for the primordial baryon asymmetry.  To generate a net
baryon asymmetry, we first show that this model generates a net lepton
asymmetry, which subsequently is transformed into a baryon
asymmetry via sphaleron reprocessing, as in
the original work of Fukugika and Yanagida~\cite{fy1}.  In the FY mechanism,
the CP-non-conserving out-of-equilibrium decays of a heavy right-handed
neutrino into ordinary light left-handed leptons (and
Higgses) produce a net lepton asymmetry much as the
out-of-equilibrium decays of super-heavy gauge or Higgs bosons were
originally used to generate a net baryon asymmetry~\cite{ww}.

The rate of decay of $\nu^c \rightarrow L + h$ is given by
$\Gamma_{D} \simeq(\frac {\lambda_{2}^{2}} {16\pi})
m_{\nu^{c}} = \frac {m_u^{2}} {16\pi v^{2}}~  \lambda_6^2~  {V^2 \over
\mu}$ where $\lambda_2 = {m_u \over v}$ is the up-quark Yukawa
coupling.  Decays will occur out of equilibrium if $\Gamma_D$ is
smaller than the expansion ratio of the Universe $H$ at
$T = {m_{\nu^{c}}}$ or $(\frac {\lambda_2^{2}} {16\pi})  M_P  <
m_{\nu^{c}}$.  In our case with $m_{\nu^{c}}\sim 10^{12}GeV$, this requirement
 becomes $\lambda_2^{2}~\la~ 10^{-6}$, which is
certainly true for first-generation Yukawa couplings.  The final
baryon asymmetry after sphaleron processing
will be $n_B\sim n_L\sim\epsilon n_\gamma$
 and ${n_B \over s} \sim 10^{-2}\epsilon$, where $\epsilon$ is the CP
asymmetry in the decay.  (For the masses
and couplings considered here, dimension-five operators of the form
$(\frac {\lambda_2^{2}} {M_{\nu^{c}}}) \nu^c \nu^c h h$ will not erase the
baryo
 and lepton asymmetries~\cite{fy2}.)

In a cosmological model with inflation, COBE observations of the
magnitude of the quadropole density fluctuations fix the overall
inflationary scale~\cite{eeno3,cdo2}. The apparent COBE discovery of primordial
microwave background fluctuations at the level $\frac {\delta \rho}
{\rho} \simeq 5 \times 10^{-6}$~\cite{Cobe} fixes the inflationary scale
 and hence the inflaton mass~\cite{cdo2}
 \begin{equation}
m_\eta\sim few \times 10^{11}~{\rm GeV}
\label{IV}
\end{equation}
and the
reheat temperature to be $T_R\simeq 10^8GeV$ in a generic inflationary
model.  The production of a lepton asymmetry by $\nu^c$ decay,
therefore, requires $m_{\nu^c}~\la~ m_\eta\sim few \times 10^{11}GeV$.
We note that with a slight generation
dependence in $\lambda_6, {\lambda_{6_1}^2 \over \lambda_{6_3}^2} = {1 \over
 10}$, we have $m_{\nu_1^c}\simeq 10^{11}GeV$.  The final baryon
asymmetry is now given by~\cite{cdo2}
\begin{equation}
{n_B \over s}\sim{n_L\over s}\sim ({m_\eta \over M_P})^{1 \over 2}\epsilon
\label{binf}
\end{equation}

As we noted earlier, the baryon (lepton) asymmetry will be diluted by the
entropy produced during the breaking of $SU(5)\times U(1)$ as $\Phi$ picks up
 its vev~\cite{cehno1}.
This dilution factor, $\Delta$,  is just the ratio of the entropy at the time
of
    $\Phi$
decay to the entropy at the time of inflaton decay (i.e., the entropy in eq.~
 (\ref{binf})
appropriately scaled),
\begin{equation}
\Delta = \frac {s(R_{d\Phi})}{s(R_{d\eta})}
(\frac {R_{d\Phi}}{R_{d\eta}})^3  =
\frac {{\alpha_\Phi}^{3/2} \tilde m^{9/2} M_P^{3/2}}{V^3 T_R^3}
(\frac {R_{d\Phi}}{R_{d\eta}})^3 =
\frac {V^3 m_\eta^{3/2}}{{\alpha_\Phi}^{1/2}
\tilde m^{3/2} M_P^3} \sim 10^3
\end{equation}
where we have taken the flaton decay rate to be
$\Gamma_\Phi = \alpha_\Phi
\frac {\tilde m^3}{V^2}$, $\alpha_\Phi \sim 10^{-3}$,
the susy breaking scale to be $\tilde m \sim 10^{-16}M_P$,
 $T_R = \frac {m_\eta^{3/2}}{M_P^{1/2}}$, $R_{d\eta (\Phi)}$ is
the cosmological scale factor at the time of inflaton (flaton) decay
  (see refs~\cite{eeno1,eeno3,cehno1} for further details),
\begin{eqnarray}
s(R_{d\Phi}) = [\rho(R_{d\Phi})]^{3/4} =
 \frac {{\alpha_\Phi}^{3/2} \tilde m^{9/2} M_P^{3/2}}{V^3}  \\
s(R_{d\eta}) =  [\rho(R_{d\eta})]^{3/4} = T_R^3
\end{eqnarray}
and
\begin{equation}
(\frac {R_{d\Phi}}{R_{d\eta}}) = \frac {m_\eta^2 V^2}{\alpha_\Phi^{2/3} \tilde
m
M_P^2}
\end{equation}
The final baryon-to-entropy ratio becomes
\begin{equation}
\frac {n_B}{s}  =  ({m_\eta \over M_P})^{1 \over 2}\epsilon {1 \over \Delta}
= \frac {\epsilon \alpha_\Phi^{1/2} \tilde m^{3/2} M_P^{5/2}}
{V^3 m_\eta}
\end{equation}
and we now estimate $\epsilon$.

When the conjugate neutrinos $\nu^{c}, \overline {\nu^{c}}$ decay, a difference
 between the
branching ratios $\nu^c \rightarrow L+ h, \overline {\nu^c} \rightarrow
\overline {L} + \bar h$ appears if CP is violated.  The dominant
contribution to this comes from the interference of the two diagrams
shown in the figure, namely the tree-level diagram and the one-Higgs-loop
radiative correction~\cite{fy1}: \begin{equation}
\epsilon = (\frac {9}{4\pi}) \frac {Im({\lambda_2}_{ij}
{{\lambda^\dagger}_2}_{jk}
 {{\lambda^\dagger}_2}_{kl} {\lambda_2}_{li})I(\frac {M^2_j}{M^2_i})}
{{\lambda_2}_{ij} {{\lambda^\dagger}_2}_{ij}}
\end{equation}
where
\begin{equation}
I (x) \equiv x^{1 \over 2} [ 1 + (1 + x)\ln ({x \over (1+x)}) ]
\end{equation}
For reasons that will soon become apparent, we expect the dominant
contribution to come from third-generation particles in the decays of
the first-generation $\nu^c$ and $\overline \nu^c$:
\begin{equation}
\epsilon_{13} = {9 \over 8\pi} |{\lambda_2}_{33}|^2 ({M_1 \over M_3}) \delta
\end{equation}
where $\delta$ is the CP-violating phase factor and we have assumed
that $|{\lambda_2}_{11}|,~|{\lambda_2}_{12}| \ll |{\lambda_2}_{13}|$.  Since
the
 see-saw light neutrino masses are expected to
be in the hierarchy $m_{\nu_\tau} (\propto {|{\lambda_2}_{33}|^2 \over
m_{\nu_1^
c})}
\gg m_{\nu_\mu} (\propto {|{\lambda_2}_{22}|^{2} \over m_{\nu_2^c}}) \gg
m_{\nu_??}
(\propto {|{\lambda_2}_{11}|^{2} \over m_{\nu_3^c}}) $,
 we see that $|\epsilon_{13}| \gg |\epsilon_{12}| > |\epsilon_{11}|$.  The
corresponding value of the present-day baryon-to-entropy ratio is
now
\begin{equation}
\frac {n_B}{n_\gamma} \simeq \epsilon (\frac {m_\eta}{M_P})^{1 \over 2}
 {1 \over \Delta}
\simeq \frac {9}{8\pi} |{\lambda_2}_{33}|^2 (\frac {m_{\nu_1^c}}{m_{\nu_3^c}})
 (\frac {m_\eta}{M_P})^{1 \over 2} {\delta \over \Delta}
\end{equation}
Or, in terms of the magnitude value of density fluctuations measured
by COBE, we can write
\begin{equation}
\frac {n_{B}}{n_{\gamma}} \simeq \frac {9}{80\pi} | {\lambda_2}_{33} |^{2}
 (\frac {m_{\nu_1^c}}{m_{\nu_3^c}})  \sqrt {\frac{\delta\rho}{\rho}}~
 {\delta \over \Delta}
\label{n+5}
\end{equation}

Recent theoretical, experimental and observational developments
enable us to put in numbers for most of the factors in the last
formula.  The value of $| {\lambda_2}_{33}|$ can be estimated from
 indirect
determinations ~\cite{efl} of $m_{t}$ via electroweak radiative corrections
$(m_
124_{-28}^{+26}GeV)$ or two irresponsible theorists' interpretation ~\cite{dg}
o
 one published event $(m_{t} = 131_{-11}^{+22}GeV)$, and the relation
\begin{equation}
|\lambda_{{2}_{33}} (m_{t})| = \frac {gm_{t}}{2M_{W} \sin\beta}
\end{equation}
where $g$ is the $SU(2)$ gauge coupling at the electroweak scale and $\tan\beta
= \frac {v_{2}}{v_{1}}$ is the ratio of supersymmetric Higgs vev's.  The latter
 is believed to be larger than unity, so that $\frac {1}{\sqrt {2}} <
\sin\beta$
  We therefore estimate that
\begin{equation}
\frac {g}{\sqrt{2}} \simeq |{\lambda_2}_{33} (m_{t})| \simeq g
\end{equation}
However, the effective renormalization scale in the CP-violating $\nu^{c}/
\overline \nu^{c}$ decays is much larger, and $h_{33} (m_{\nu^{c}})$
 may be somewhat smaller, so we take
\begin{equation}
h_{33} (m_{\nu^{c}}) \simeq \frac {g} {\sqrt{2}}
\end{equation}
as our central value.  This coincides with the value calculated form
first principles in versions of flipped $SU(5)$ derived from string theory,
assuming plausible mixing between the $t$ quark and other charge
$\frac {2}{3}$ fields.

The values of the ratios $(\frac {M_{1}}{M_{2,3}})$ were discussed
 in ref.~\cite{eln}.  We
estimated there that $(\frac {m_{\nu_{2}^{c}}} {m_{\nu_{1}^{c}}}) = 0(1)$ on
the
 basis of
 the known values of $m_{u}$ and $m_{c}$ and the value of the $\nu_{e} -
\nu_{\mu}$ mixing angle that best fits the solar neutrino data, assuming
 the MSW mechanism.
We also found that the best fit to the combination of COBE and other
data on primordial density perturbations, which prefer an admixture
of about 30\% hot dark matter for which the best (only?) candidate is
a $\nu_{\tau}$ weighing $0(10)eV$, would be with $\frac {m_{\nu_{3}^{c}}}
 {m_{\nu_{2}^{c}}} = 0(10)$.  Therefore we put $\frac {m_{\nu_{1}^{c}}}
{m_{\nu_{3}^{c}}}\simeq \frac{1}{10}$ in equation (\ref{n+5}).

The only remaining unknown is the CP-violating phase factor $\delta$
 in (\ref{n+5}) and we estimate
\begin{equation}
 \frac {n_{B}}{n_{\gamma}} \simeq 2 \times 10^{-6} {\delta \over \Delta}
\simeq  2 \times 10^{-9} \delta
\end{equation}
We conclude that for plausible values of $\delta$, this mechanism is
completely consistent with the value $\frac {n_{B}}{n_{\gamma}} \sim 3 \times
10^{-10}$ extracted from the
concordance of data on primordial nucleosynthesis~\cite{wssok}, even with the
possibility of some extra entropy generation subsequent to the
 $\nu^{c}$ and $\overline \nu^{c}$ decays.

We find it remarkable that the simplest flipped
SU(5) model, in
addition to all its other well-documented virtues and its motivation
from string theory, also ties together in a very economical way
baryogenesis, solar neutrino physics, COBE and Dark Matter. The same
flipped heavy neutrino that plays a key role in the see-saw mass matrix,
accommodating the MSW interpretation of the apparent solar neutrino
deficit and the apparent preference for an admixture of hot Dark Matter,
also generates a lepton (and hence, thanks to sphalerons, a baryon)
asymmetry in a very simple and convincing way. Many of the parameters
needed to make a quantitative estimate of the resulting baryon-to-entropy
ratio have recently been fixed or severely constrained by
accelerator, GALLEX and COBE data. Furthermore, as we have emphasized
previously~\cite{eln}, these astrophysical aspects of the flipped SU(5) model
ca
in principle soon be tested via accelerator experiments searching for
$\nu_{\mu}$ - $\nu_{\tau}$ oscillations.}}

\noindent{ {\bf Acknowledgements} } \\
\noindent The work of DVN was supported in part by DOE grant
DE-FG05-91-ER-40633 and by a grant from Conoco Inc. The work
of KAO was supported in part by DOE grant DE-AC02-83ER-40105, and by
a Presidential Young Investigator Award.

\newpage

\noindent
{\bf Figure caption}
\vskip 0.5cm
\noindent
The tree and one-Higgs-loop diagrams whose interference gives the
largest contribution to the lepton asymmetry in heavy singlet
neutrino decay in flipped SU(5).
\vglue 6.5cm
\hglue 1.5 cm
\special{picture fig1 scaled 900}

\end{document}